\documentclass[
    ,final            
  ]
  {aipproc}

\layoutstyle{6x9}

\def\gtap{\mathrel{\hbox{\rlap{\lower.55ex \hbox {$\sim$}}
                   \kern-.3em \raise.4ex \hbox{$>$}}}}
\def\ltap{\mathrel{\hbox{\rlap{\lower.55ex \hbox {$\sim$}}
                 \kern-.3em \raise.4ex \hbox{$<$}}}}

\begin{document}

\title{Creating ultra-compact binaries through stable mass transfer}


\author{M.V. van der Sluys}{
  address={Astronomical Institute, Postbox 80.000, 3508 TA Utrecht, the Netherlands}
}

\author{F. Verbunt}{
  address={Astronomical Institute, Postbox 80.000, 3508 TA Utrecht, the Netherlands}
}

\author{O.R. Pols}{
  address={Astronomical Institute, Postbox 80.000, 3508 TA Utrecht, the Netherlands}
}

\begin{abstract}
A binary in which a slightly evolved star starts mass transfer to a neutron star can evolve towards
ultra-short orbital periods under the influence of magnetic braking.  This is called magnetic capture.  
We investigate in detail for which initial orbital periods and initial 
donor masses binaries evolve to periods less than 30--40 minutes within the Hubble time.  
We show that only small ranges of initial periods and masses lead to ultra-short periods, and that
for those only a small time interval is spent at ultra-short periods.
Consequently, only a very small fraction of any population of X-ray 
binaries is expected to be observed at ultra-short period at any time.  If 2 to 6 of the 13 bright X-ray
sources in globular clusters have an ultra-short period, as suggested by recent observations, their 
formation cannot be explained by the magnetic capture model.
\end{abstract}

\maketitle


\section{Introduction}
  About half of the bright X-ray sources in the galactic globular clusters possibly are 
  binaries with ultra-short orbital periods ($\ltap 40$\,min).  Two of the five periods known 
  are 11.4\,min (in NGC\,6624) and 20.6\,min (in NGC\,6712). The 11.4\,min system has a negative 
  period derivative.
  This high fraction of ultra-short periods is in marked contrast to bright X-ray sources in 
  the galactic disk, where such periods are less common (See Table~3 of \cite{cc04}).
  
  One of the scenarios to explain the ultra-short periods starts from a binary with 
  a neutron star and a main-sequence star.  For a small range of initial orbital
  periods, strong magnetic braking can shrink the orbit sufficiently that
  the system evolves to a minimum period in the ultra-short range. This way, an orbital 
  period shorter  than 11\,min can be reached \cite{podsiadlowski}.  At 11.4\,min, 
  the period derivative may be negative or positive, depending on whether the system 
  evolves to the period minimum, or has already rebounded. 
  
  We will try to find out which initial systems can reach ultra-short periods within
  the age of the globular clusters and what the chances are to observe these systems
  as X-ray sources.  This research is published as an article in \cite{sluys}.

\section*{The evolution code}
  We calculate our models using the \textsc{STARS} binary stellar evolution
  code, developed by Eggleton \cite{eggleton}, but with updated physics \cite{pols}.  
  The primary is treated as a point mass with an initial mass of $1.4\,M_\odot$.  Sources of angular momentum loss are
  gravitational waves, partially conservative mass transfer, and magnetic braking according to \cite{rappaport}:
  \begin{equation}
    \frac{dJ_{\rm MB}}{dt} ~=~ -3.8 \times 10^{-30} \, M_2 \, R_2^4 \, \omega^3 \, {\rm dyn~cm} .
    \label{eq:magnbrak}
  \end{equation}
  where $M_2$, $R_2$ and $\omega$ are the mass, radius and angular rotational velocity of the secondary respectively.
  Tidal effects keep the spin synchronised to the orbit and magnetic braking effectively removes 
  angular momentum from the orbit.

\section*{Results}
  We have calculated a grid of models for $Z=0.01$ (the metallicity of NGC\,6624), 
  with initial masses between 0.7 and 1.5\,$M_\odot$ and initial periods
  between 0.35 and 3\,d.  The grid is refined in period around the bifurcation period 
  between converging and diverging systems.  We consider models that converge after the
  Hubble time as diverging.  

  \begin{figure}   
    \includegraphics[angle=-90,width=12.3cm]{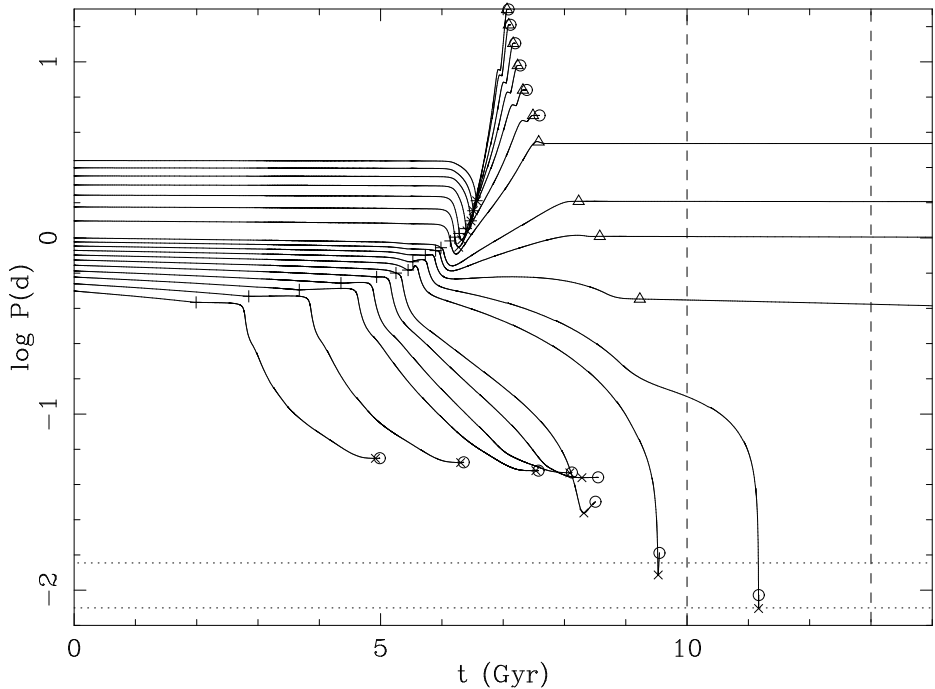}  
    \caption{Time-period (t-P) tracks for $1.1\,M_\odot$. Initial periods are spaced 0.05\,d
      below 1\,d and 0.25\,d above that.
      The symbols show: start of mass transfer +; period minimum $\times$;
      end of mass transfer $\bigtriangleup$; the the last model {\tt o}. The dotted 
      lines are at 11.4 and 20.6\,min.} 
    \label{fig:tp}
  \end{figure}   

  Figure\,\ref{fig:tp} shows that orbits with low initial period converge to about 70\,min, and orbits with
  high orbital period diverge to several days.  A small range in between leads to ultra-compact
  systems. It is clear that an initial period must be picked carefully to find such a short
  period minimum.

\begin{figure}   
    \includegraphics[angle=-90,width=12.3cm]{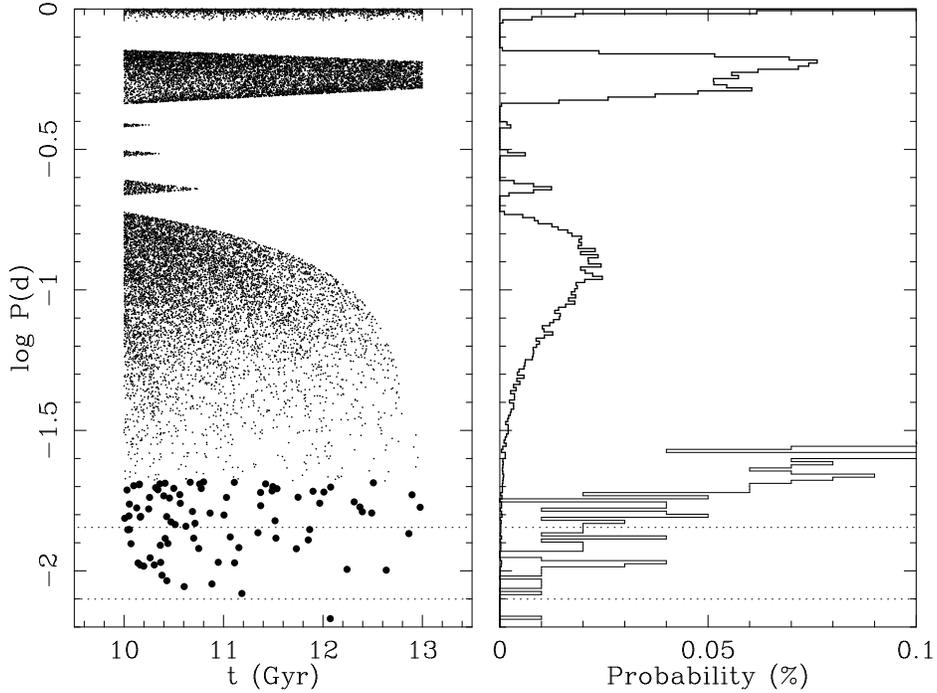} 
    \caption{Left panel (a): accepted systems for 1.1$M_\odot$. Each dot represents one system.
      The spikes at log$P\approx-0.5$ are artefacts due to the interpolation. Dots below 30\,min are 
      larger for more clarity. Right panel (b): Histogram of the data
      obtained by summing (a) over the time. The thin line shows the short-period tail, with the probability 
      100 times enlarged.} 
    \label{fig:phist}
  \end{figure}

  To determine the probability of observing an ultra-compact binary produced this way, we perform 
  statistics on the t-P tracks. We choose a random initial period from a flat distribution in $\log P$ and
  determine the t-P track of a system with that initial period by interpolation.  Once the track is known, we choose a
  random moment in time between 10 and 13\,Gyr (the dashed lines in Fig.\,\ref{fig:tp}) and determine the orbital 
  period at that moment.  If a system has passed its period minimum, or has no mass transfer at
  that moment, we reject it.  For $1.1\,M_\odot$, 10.5\% of the $10^6$ probes is accepted and shown in Fig.\,\ref{fig:phist}a. \
  Of these, 86 systems have an orbital period less than 30\,min (the large dots in the figure).
  Figure\,\ref{fig:phist}b shows the corresponding period distribution.

\begin{figure}   
    \includegraphics[angle=-90,width=12.3cm]{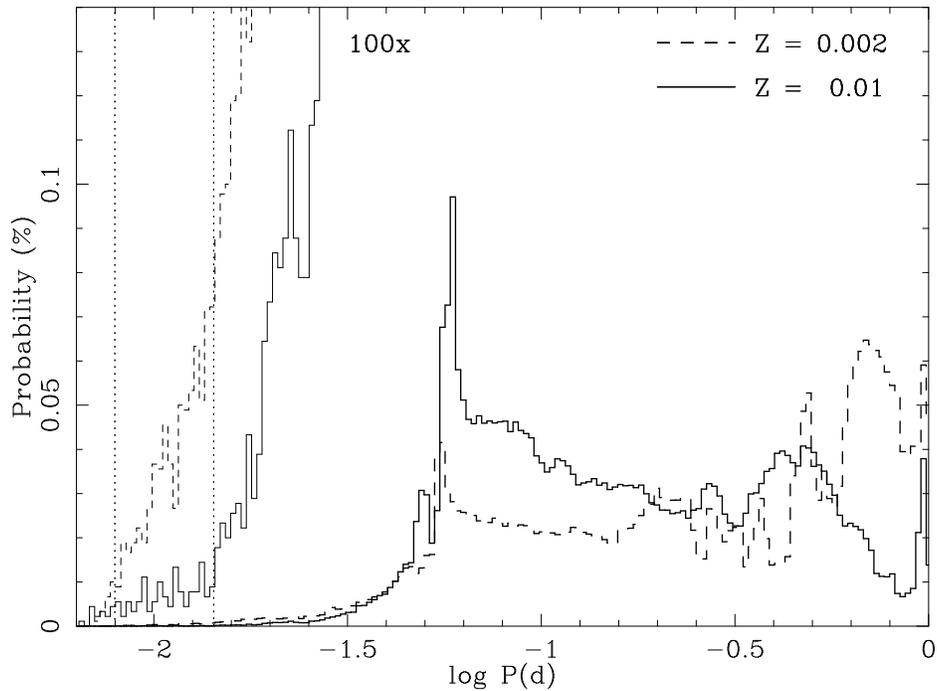} 
    \caption{Period distributions for $Z=0.002$ and $Z=0.01$.  The thin lines
      on the left are enlarged 100 times in the vertical direction.} 
    \label{fig:allz}
  \end{figure}

   We have calculated similar grids for $Z=0.002$ and $Z=0.02$ and performed the same
   statistics as for $Z=0.01$.  The period distributions for each mass have been added
   using a flat mass distribution. A Salpeter mass distribution leads to little 
   difference, especially for the ultra-short period regime.  We show the results for the
   two of these metallicities in Fig.\,\ref{fig:allz}.

   Figure\,\ref{fig:allz} shows firstly that there are no big differences in the expected fraction of observable
   ultra-compact X-ray binaries.  Though the exact numbers are uncertain due to the low number
   of accepted ultra-compact systems, we expect that of a population of $10^7$ binaries with 
   initial periods between roughly 0.5 and 2.5\,d, 1 to 10 systems have a period of 11.4\,min and 
   10 to 100 systems have a 20.6\,min orbital period and emit X-rays.

  \section*{Discussion}
  We confirm that magnetic capture can lead to ultra-compact X-ray binaries within a
  Hubble time.  In order to find a binary system that will have its minimum period in the
  ultra-short period regime, one has to carefully select an initial period, just under the
  bifurcation period for that system.  The systems that reach an ultra-short period 
  remain there for a relatively short time, as can be seen from the steep tracks
  in Fig.\,1.  These factors combined make it rather unlikely that such a system can be 
  observed.  The metallicity of the stars only has a small influence.
  
  Alternative scenarios to create ultra-compact binaries only allow positive period
  derivatives for these binaries. For a description and references see Verbunt (these proceedings)
  and \cite{sluys}.




\bibliographystyle{aipproc}   

\bibliography{sluys}

\begin{thebibliography}{6}
\expandafter\ifx\csname natexlab\endcsname\relax\def\natexlab#1{#1}\fi
\providecommand{\enquote}[1]{``#1''}
\expandafter\ifx\csname url\endcsname\relax
  \def\url#1{\texttt{#1}}\fi
\expandafter\ifx\csname urlprefix\endcsname\relax\def\urlprefix{URL }\fi
\providecommand{\eprint}[2][]{\url{#2}}

\bibitem[Charles and Coe(2004)]{cc04}
P.~Charles, and M.~Coe, \enquote{A catalogue of low-mass X-ray binaries,} in
  \emph{Compact stellar X-ray sources}, edited by W.~Lewin, and M.~van~der
  Klis, Cambridge U.P., Cambridge, 2004.

\bibitem[{Podsiadlowski} et~al.(2002)]{podsiadlowski}
P.~{Podsiadlowski}, S.~{Rappaport}, and E.~D. {Pfahl}, \emph{ApJ},
  \textbf{565}, 1107--1133 (2002).

\bibitem[{Van der Sluys} et~al.(2004)]{sluys}
M.~{Van der Sluys}, F.~{Verbunt}, and O.~{Pols}, \textbf{astro{-}ph{/}0411189},
  {\it A\&A}, in press (2004).

\bibitem[{Eggleton} and {Kiseleva-Eggleton}(2002)]{eggleton}
P.~P. {Eggleton}, and L.~{Kiseleva-Eggleton}, \emph{ApJ}, \textbf{575},
  461--473 (2002).

\bibitem[{Pols} et~al.(1995)]{pols}
O.~R. {Pols}, C.~A. {Tout}, P.~P. {Eggleton}, and Z.~{Han}, \emph{MNRAS},
  \textbf{274}, 964--974 (1995).

\bibitem[{Rappaport} et~al.(1983)]{rappaport}
S.~{Rappaport}, P.~C. {Joss}, and F.~{Verbunt}, \emph{ApJ}, \textbf{275},
  713--731 (1983).

\end{thebibliography}

\end{document}